# Vacuum squeezing enhanced micrometer scale vapor cell magnetometer

Shahar Monsa, Yair Chasid, Michael Shuldiner, Shmuel Sternklar, Eliran Talker*

Department of Electrical and Electronic Engineering, Ariel University, Ariel 40700, Israel

Corresponding author: elirant@ariel.ac.il

We report on an optical magnetometer enhanced by vacuum-squeezed light, employing an Mx magnetometer based on $^{87}$Rb vapor in a micrometer-scale cell (~100 μm). Using the well-established polarization self-rotation effect in a room-temperature $^{87}$Rb vapor cell, we achieve -3 dB of vacuum squeezing within the noise spectral window of 100 Hz to several MHz, corresponding to 3.5 dB squeezing when accounting for optical losses. Leveraging this level of squeezing, we demonstrate a magnetic field sensitivity of approx. 1 pT/√Hz. The combination of vacuum-squeezed light and micrometer-scale vapor cells paves the way for compact, low-power-consumption atomic sensors with enhanced performance.

Introduction: Optical magnetometers are currently the most sensitive devices for measuring low-frequency magnetic fields [1–3], with applications ranging from medical diagnostics and biomagnetism to space exploration and tests of fundamental physics. While enhanced sensitivity due to squeezed light has been demonstrated in centimeter-scale vapor cells [4–7], no such demonstration has yet been achieved in micrometer-scale vapor cells. In micrometer-scale vapor cells, the signal-to-noise ratio (SNR) is inherently lower than in their centimeter-scale counterpart due to the lower optical density (OD). Therefore, employing optical squeezed light can enhance the SNR and, consequently, improve the sensitivity of optical magnetometers in these miniaturized systems. In this paper, we demonstrate a squeezed-light-enhanced optical magnetometer achieving a detection noise level of approximately 1 pT/√Hz, using our homemade micrometer-scale (~100 μm) vapor cell. Vacuum squeezed light is generated via the polarization self-rotation (PSR) effect [8–14]. In this technique, both the vacuum squeezed light and the local oscillator beams interact within the vapor cell for angle rotation measurements and external magnetic field sensing. Our magnetometer operates in the Mx configuration [15–21], where the external magnetic field $B_0$ to be measured is oriented at 45° with respect to the laser beam, and an oscillating magnetic field ($B_\omega$), with a frequency equal to the Larmor frequency, is oriented perpendicular to the direction of the light beam. Classically, the Larmor precession around $B_0$ is driven by the co-rotating component of the oscillating field, which imparts a phase shift on the precessing spins. The projection of this precessing polarization onto the direction of the light beam leads to an oscillating magnetization along the beam axis, resulting in periodic modulation of the optical absorption coefficient. The sensitivity of an Mx magnetometer is fundamentally limited by two main noise sources: atomic projection noise and optical shot noise. When atomic projection noise dominates, sensitivity can be improved using quantum non-demolition (QND) measurements [22], atomic entanglement [23], and spin squeezing [24,25], particularly within the atomic coherence time and for non-exponential relaxation processes. Conversely, when optical shot noise is the limiting factor, as in the case in our micrometer-scale cell due to its short propagation length, optical squeezing becomes an effective method for noise reduction. Therefore, our focus in this work is to reduce system noise using squeezed light. This paper is organized as follows: we first present the experimental setup, followed by a discussion of the results, and conclude with a summary of our findings. The demonstration of micrometer-scale cell magnetometry with high sensitivity forms the basis for developing compact, low-power, and highly precise magnetic sensors, enabling applications ranging from biomedical diagnostics and brain imaging to portable navigation devices and fundamental physics experiments.

Experimental setup and results: The schematic of the experiment is shown in FIG. 1. A Toptica DL pro extended cavity diode laser with an output power of 100 mW is stabilized to 350 MHz red-detuned from the [87]Rb D1 $5^2S_{1/2} \to 5^2P_{1/2}$ (F = 2 → F′ = 2) transition using the well-known saturated absorption spectroscopy (SAS) technique [26,27] (FIG. 1(a)). Our micrometer-scale vapor cell contains [87]Rb, 250 Torr of nitrogen ($N_2$), and 350 Torr of helium (He). The buffer gases serve to suppress spin depolarization caused by wall collisions, with $N_2$ also acting as a quenching gas. The specific ratio of $N_2$ to He is optimized to minimize energy level shifts. The influence of the buffer gas on the transmission spectra is depicted in FIG 1(b), which shows the measured transmission signal from the micrometer-scale cell at ∼100 °C, compared with a 7.5 cm reference cell held at room temperature (which does not contain buffer gases). The generation of the vacuum squeezed light is illustrated in FIG. 2. The laser beam is first linearly polarized using a $\lambda/2$ plate followed by a polarizing beam splitter (PBS). This allows for fine control over the laser beam power, and enhancement of the polarization purity of the incident light. The beam intensity is set to 10 mW and then focused into a 75 mm-long Rb vapor cell (FIG. 2(a)). The cell is surrounded by three layers of magnetic shielding with end caps to eliminate ambient magnetic fields and ensure a near-zero-field environment. We also use a set of Helmholtz coils to actively cancel any stray magnetic field. The cell is maintained at 50 °C, corresponding to an atomic number density of $1.1 \times 10^{11}$ atoms/cm³. To evaluate the degree of squeezing, we employed a flip mirror to direct the beam into a balanced polarimeter setup, which consisted of a $\lambda/2$-wave plate, a PBS, and two balanced photodetectors. The $\lambda/2$-wave plate was adjusted to balance the powers in the TE and TM linearly polarized beams at the outputs of the PBS when the frequency of the laser was tuned far from resonance. To measure the quadrature noise properties of the TE linearly polarized vacuum beam, we then performed homodyne detection, as shown in FIG 1, using the TM linearly polarized output of the PBS as a local oscillator (see FIG. 2(b)). The squeezed vacuum beam along with local oscillator beam (which for the magnetometer measurements fulfills the role of the probe beam) are then used to implement an Mx magnetometer as illustrated in FIG. 2(c). The core of the Mx magnetometer is a custom-fabricated micrometer-scale vapor cell where the probe beam is focused to a waist of 900 μm (see FIG. 1(d)). The complete fabrication process is detailed in [28–30]. Here, we briefly summarize the key steps. The cell is constructed using anodic bonding between glass and an amorphous silicon substrate, performed at a temperature of 300 °C and an applied voltage of 800 V. Following the bonding process, rubidium is introduced into the sealed microcell using a distillation method, along with controlled pressures of buffer gases. For all Mx magnetometer measurements, the micrometer cell is maintained at 100 °C, corresponding to an atomic density of $1.8 \times 10^{12}$ atoms/cm³. The magnetometer operates by measuring the Faraday rotation of the probe beam. The micrometer cell is housed inside a four-layer magnetic shield to suppress environmental magnetic noise. Additionally, the shielding contains two sets of Helmholtz coils: one to generate the static magnetic field under investigation, and another to apply the RF field required for the Mx magnetometer operation (only one set is depicted in FIG 2(c)). The Faraday rotation angle is measured using a polarimeter setup consisting of a half-wave plate, PBS, and a balanced photodetector. The signal is demodulated using a lock-in amplifier referenced to the Larmor frequency, and the resulting magnetic resonance signals are recorded and analyzed on a PC. FIG. 3 shows the measured noise oscillations obtained by scanning the phase of the coherent local oscillator using a piezoelectric actuator and recording the signal with a balanced detector. We achieve a maximum measured squeezing of 3 ± 0.25 dB. After correcting for linear losses, photodiode inefficiencies, and electronic noise of the photodetector, this corresponds to an inferred squeezing level of 3.5 ± 0.35 dB. To confirm that the degree of squeezing remains almost the same after the Mx magnetometer Rb cell, we characterized the squeezing statistics while carefully accounting for all system losses, including those introduced by the second magnetometer cell, the quarter-wave plate, and the polarizing beam splitter. Remarkably, we were able to measure nearly the same degree of squeezing as initially generated, demonstrating that our setup effectively preserves the quantum noise reduction even after passing through the additional magnetometer stage. This low squeezing loss can be attributed to the very low optical absorption in the micrometer-scale cell, which minimizes losses and preserves the degree of squeezing. We emphasize that using a low optical density (OD) system — achievable by reducing the vapor cell length to the micrometer scale — greatly minimizes absorption and thereby preserves the degree of vacuum squeezing much more effectively than traditional millimeter-scale vapor cells. In our system, polarization-squeezed light is described using quantum Stokes operators, which capture fluctuations in different polarization components. When light interacts with a nonlinear Kerr medium, it can develop polarization squeezing through self- and cross-phase modulation terms. The achievable squeezing depends on the interaction strength and relative phases and is ultimately quantified by the variance reduction in certain Stokes parameters. Crucially, vacuum-squeezed light is very sensitive to losses from absorption and scattering. Such losses can be modeled as an effective beamsplitter with transmissivity η, where the final measurable squeezing $S_m$ degrades as [31]

$$S_m = \eta S_{in} + (1 - \eta) \qquad (1)$$

Where $S_{in}$ is the initial squeezing and it is equal [32]

$$S = \frac{V_j}{|\langle \hat{S}_1 \rangle|} = \frac{\bar{n}_h + \bar{n}_v - 2|\gamma_h - \gamma_v|\bar{n}_h \bar{n}_v}{|\bar{n}_h + \bar{n}_v|} \qquad (2)$$

Where $\bar{n}_h, \bar{n}_v$ is the number of horizontal and vertical photon number, and $\gamma_{j=\{h,v\}}$ is proportional to $\chi^{(3)}$. $\hat{S}_1$ is the first Stokes operator [32]. Using ultra-thin cells with lower OD, absorption losses are strongly suppressed (higher η), resulting in minimal squeezing degradation. Thus, shortening the vapor cell length directly reduces losses and significantly improves the preservation of quantum noise reduction, highlighting a major advantage of micrometer-scale vapor cells in quantum optics applications. Next, both the squeezed vacuum beam and the local oscillator beam (i.e. the probe beam) are used to perform Mx magnetometry. The magnetic field to be measured ($B_0$ = 5.22 μT) is provided by a set of Helmholtz coils whose axis is oriented 45°C with respect to the optical beam. A second set of Helmholtz coils produces the RF field to coherently drive the precession of the atomic spin about the static magnetic field. The $^{87}$Rb atomic spin precession modulates the transmitted light intensity, which is collected by a balanced photodetector. The RF light current is adjusted to optimize the sensitivity of the magnetometer. FIG 4 shows the typical resonance line observed by demodulating the signal with a lock-in amplifier. FIG. 4(a) shows the demodulated spectra obtained from the vacuum squeezed state light compared to a coherent probe light (FIG. 4(b)). The sensitivity ($\delta B$) of the magnetometer, defined as the minimum detectable change in magnetic field, is given by [20]:

$$\delta B = \frac{1}{\gamma} \frac{\Delta \nu}{SNR} \quad (3)$$

where $\Delta \nu$ is the magnetic resonance linewidth (measured as half-width at half-maximum), and $\gamma = 4.7 \ Hz/nT$ is the gyromagnetic ratio of $^{87}$Rb. Due to an improved SNR by a factor of approx. 3, we expect a corresponding threefold enhancement in magnetic sensitivity., This expectation is confirmed experimentally as shown in FIG. 5: the magnetic noise sensitivity is improved from ~3 $pT/\sqrt{Hz}$ to ~1 $pT/\sqrt{Hz}$.

In conclusion, we have demonstrated a significant advancement in optical magnetometry by integrating vacuum-squeezed light with a micrometer-scale 87Rb vapor cell. Utilizing the PSR effect in room-temperature vapor, we achieved -3 dB of vacuum squeezing, or an effective -4 dB after accounting for optical loss. This enhanced quantum resource enabled us to achieve a magnetic field sensitivity of approximately 1 $pT/\sqrt{Hz}$, a performance level that is typically challenging for sensors operating with such a small interaction volume. Our results clearly show that the combination of squeezed light with micrometer-scale vapor cells not only compensates for the reduced optical path length inherent to miniaturized systems but also significantly boosts sensitivity. This approach paves the way for a new class of compact, low-power, quantum-enhanced atomic sensors, with potential applications in portable magnetometry, biomagnetic field detection, navigation, and other areas where size, weight, and power consumption are critical constraints.

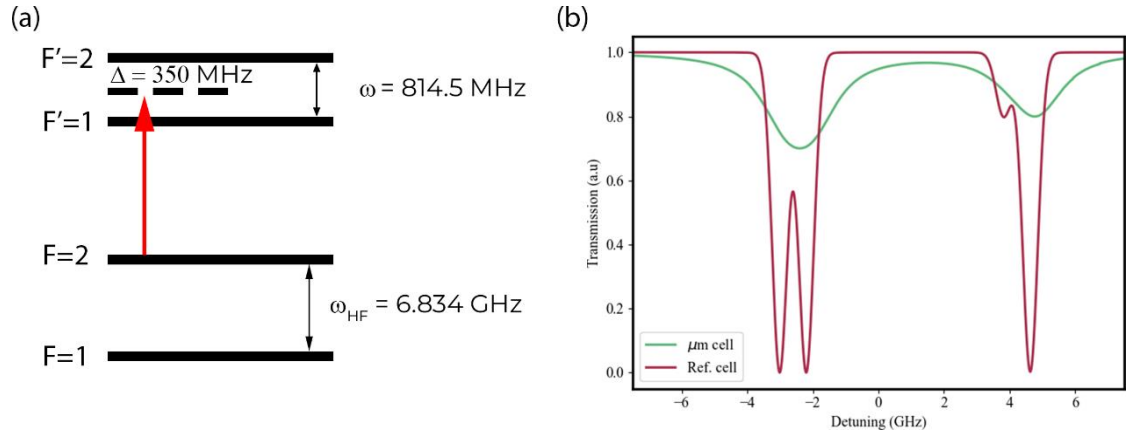

**FIG. 1**: (a) Energy-level diagram of the D1 transition of $^{87}$Rb; (b) Transmission spectroscopy of the micrometer cell (green line) and reference cell (red line), where the laser beam intensity is 500 μW, the reference cell is at room temperature and the micrometer cell temperature is set to ~100°C

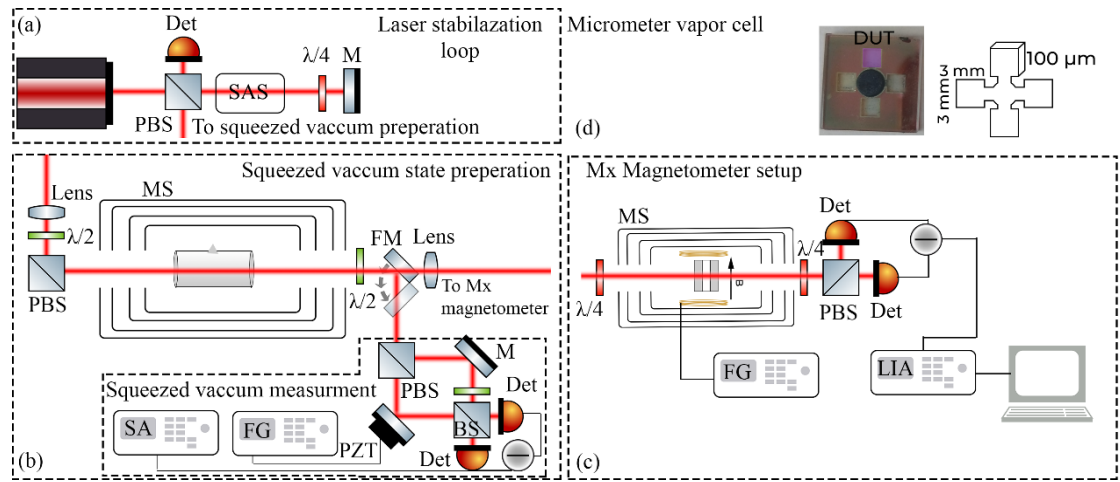

**FIG. 2** (a) Laser stabilization loop in which a small portion of the beam enters the Rb vapor cell (7.5 cm) and back reflected to the detector to observe the Doppler free spectra. The laser is frequency-stabilized around the $^{87}$Rb $F = 2 \rightarrow F' = 2$. (b) System for squeezed vacuum-state preparation. (c) Mx-magnetometer setup, where a circularly polarized light beam is directed through a micrometer-scale vapor cell. By applying an oscillating magnetic field perpendicular to the light propagation direction, we induce Larmor precession of the atomic spins. The resulting Faraday rotation of the light polarization, caused by the spin dynamics, is detected using a polarimetric scheme, enabling precise measurement of the magnetic field. (d) image and sketch of the micrometer cell. MS – magnetic shielding, PBS-polarizing beam splitter, FM – flip mirror, Det – detector, M – mirror, FG – function generator, SA – spectrum analyzer, LIA – lock-in amplifier, SAS – saturated absorption spectroscopy, BS – beam splitter, $\lambda/4$ – quarter-wave plate, $\lambda/2$ – half-wave plate. DUT – device under test.

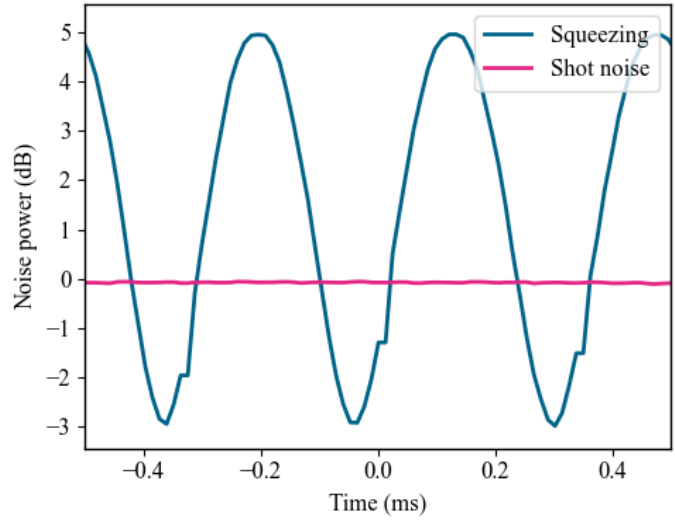

**FIG. 3.** Polarization noise power measured by the detector at a center frequency of 100 kHz in zero-span mode, with a resolution bandwidth of 10 kHz and a video bandwidth of 100 Hz. The pink trace indicates the photon shot noise level of a coherent probe, taken as a reference at 0 dB. The blue trace shows the quantum noise oscillations below (squeezing) and above (anti-squeezing) the shot noise level, acquired while scanning the local oscillator phase.

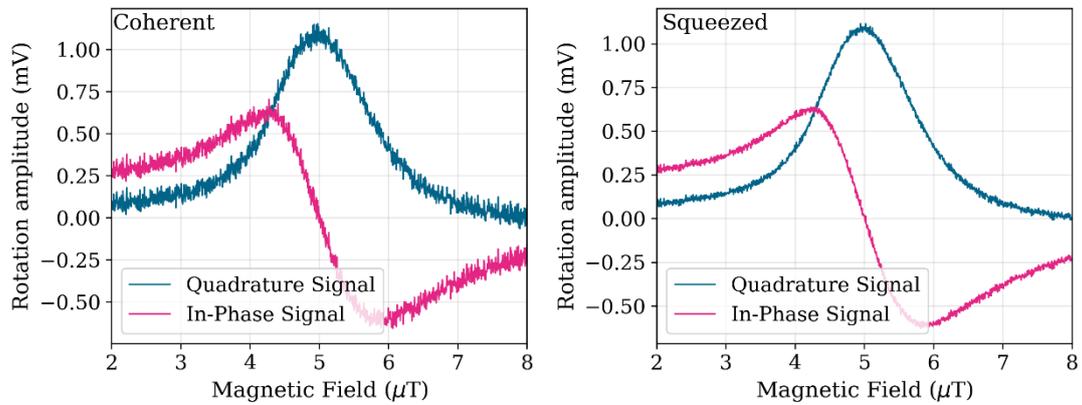

**FIG. 4.** Magnetic resonance spectra obtained by scanning the frequency of the oscillating field for (a) vacuum squeezed light (b) coherent light. blue line - quadrature component, pink line - in-phase component.

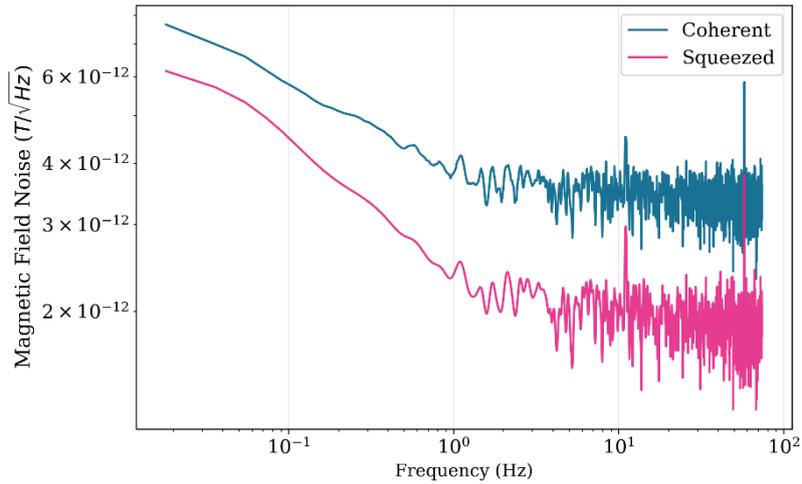

**FIG. 5**. Magnetic sensitivity measurement showing the measured magnetic sensitivity (blue) for the case of coherent light source, and for the squeezed vacuum light source (red). The probe beam intensity was set to 10 mW and the cell temperature is approx. $100°C$.